\documentclass[useAMS,usenatbib]{mn2e}
\usepackage[T1]{fontenc}
\usepackage{aecompl}
\usepackage{graphicx}

\title[Early X-ray binaries]{X-Ray Luminous Binaries, Metallicity, and the Early Universe}

\author[P.\ Kaaret]{Philip Kaaret\\
Department of Physics and Astronomy, University of Iowa, Iowa City, IA 52242}

\begin{document}

\date{Accepted 2014 January 22. Received 2014 January 12; in original form 2013 December 09}

\pagerange{\pageref{firstpage}--\pageref{lastpage}} \pubyear{2014}

\maketitle

\label{firstpage}

\begin{abstract}

High mass X-ray binaries (HMXBs) may have had a significant impact on the heating of the intergalactic medium in the early universe.  Study of HMXBs in nearby, low metallicity galaxies that are local analogues to early galaxies can help us understand early HMXBs.  The total luminosity of HMXB populations is dominated by sources at high luminosities.  These sources exhibit X-ray spectra that show curvature above 2~keV and the same is likely true of HMXB populations at high redshifts.  The spectral curvature changes the K-correction for X-rays from HMXBs in a manner that weakens the constraints on X-ray emission of early HMXBs obtained from the soft X-ray background.  Applied to deep X-ray surveys of star forming galaxies, the modified K-correction suggests a moderate increase in the ratio of X-ray luminosity to star formation rate at intermediate redshifts, $z=3-5$, and is consistent with a large enhancement at high redshifts, $z=6-7$.

\end{abstract}


\section{Introduction}

Between 10$^{8}$ and 10$^{9}$ years after the Big Bang, the intergalactic medium (IGM) changed from being cold and neutral to being warm and ionized.  An early generation of massive stars with very low metal content producing copious UV radiation is the most likely cause of reionization \citep{Loeb10}.  X-rays from accretion onto compact objects, either stellar mass or supermassive, are thought to have heated the IGM and may have made a $\sim 10$ per cent contribution to reionization \citep{Venkatesan01,Ricotti04}.  X-rays from active galactic nuclei (AGN) dominate the X-ray background at $z < 5$, but recent work places strong constraints on X-ray emission from AGN at high $z$ \citep{Salvaterra12,Treister13}.  Recent work on the formation of the first stars suggest that they had typical masses of a few tens of $M_{\odot}$ and formed in small multiple systems often dominated by a binary \citep{Bromm13}.  The formation of supermassive objects, $\ge 10^{5} M_{\odot}$, was rare \citep{Bromm13}.  Thus, accretion onto stellar-mass compact objects likely dominated the X-ray emission in the early universe with a significant fraction occurring in binary systems \citep{Jeon13}.


It is not possible to detect HMXBs in the early universe using current instruments to obtain direct observational constraints.  A number of authors have considered the effects of HMXBS on the early universe with inferences about their properties drawn from theory or observations of Milky Way black hole binaries and have assumed hard power-law spectra with photon indices of 1.3--2.0 \citep{Madau99,Venkatesan01,Pritchard07,Mirabel11,Wheeler11,Jeon13} sometimes with a soft thermal component \citep{Ricotti04,Power13,Fragos13} providing a minority of the total energy output.  Here, we attempt to infer the properties of early HMXBs based on local analogues in nearby galaxies.  The early universe was highly metal deficient \citep{Loeb10}.  Metallicity appears to have a strong impact on the formation and properties of X-ray binaries.  We discuss the properties of X-ray binaries found in low metallicity environments in the local universe in Section~\ref{sec:binaries} and some of the implications for the study of binaries in the early universe in Section~\ref{sec:implications}.  Section~\ref{sec:summary} is a brief summary.


\section{X-ray binaries at low metallicity}
\label{sec:binaries}

The bulk of X-ray emission from galaxies not hosting an active nucleus arises from compact objects formed at the endpoint of the stellar evolution of massive stars (typically $> 8 M_{\odot}$) in binary systems.  The X-ray binaries that form within 100~Myr after birth of their parent stars are dominated by high mass X-ray binaries (HMXBs) in which the companion star has a high mass and therefore evolves rapidly.  HMXBs also dominated the X-ray binary population of the early universe due to their rapid evolution.  The number of HMXBs and the total X-ray luminosity they produce is well correlated with the host galaxy star formation rate \citep{Griffiths90,David92,Ranalli03,Grimm03,Kaaret08}.  Furthermore, the occurrence of very bright HMXBs or ultraluminous X-ray sources (ULXs), defined here as having luminosities above $10^{39} \rm \, erg \, s^{-1}$, is correlated with recent starburst activity \citep{Irwin04}.

The number of ULXs in \citep{Mapelli10} and the total X-ray luminosity of \citep{Basu13b} a galaxy, normalised by its star formation rate (SFR), increases with decreasing metallicity.  These trends are enhanced at very low metalliticies, $Z/Z_{\odot} < 0.1$, where the number of ULXs increases by a factor of $7 \pm 3$ (normalized to SFR) relative to near-solar-metallicity galaxies \citep{Prestwich13} and the total X-ray luminosity increases by a factor of $9.7 \pm 3.2$ \citep{Kaaret11,Brorby14}.

Population synthesis simulations of the formation and evolution of high mass X-ray binaries in low metallicity environments also suggest enhanced X-ray emission.  \citet{Dray06} found that the number of HMXBs (per unit SFR) increased about a factor of 3 between solar and Small Magellanic Cloud (SMC) metallicity $Z/Z_{\odot} = 0.2$.  \citet{Linden10} found that the number of HMXBs at $Z/Z_{\odot} = 0.02$ increased by a factor of $\sim 3.5$ relative to solar metallicity and the number of ULXs increased by a factor of $\sim 5$ due to an increase in the number of binary systems evolving through the pathways that form Roche-lobe overflow systems and supergiant wind accretors. Also, the maximum black hole mass increases with decreasing metallicity \citep{Fryer99,Fryer01,Heger03,Mapelli09,Zampieri09}, reaching as high as $80 M_{\odot}$ for $Z/Z_{\odot} = 0.01$ compared with $\sim 20 M_{\odot}$ for $Z/Z_{\odot} \sim 1$ \citep{Belczynski10}.  Thus, it is possible to form more luminous X-ray binaries at low metallicity without violating the Eddington limit.

\begin{figure}
\centerline{\includegraphics[width=3.2in,angle=0]{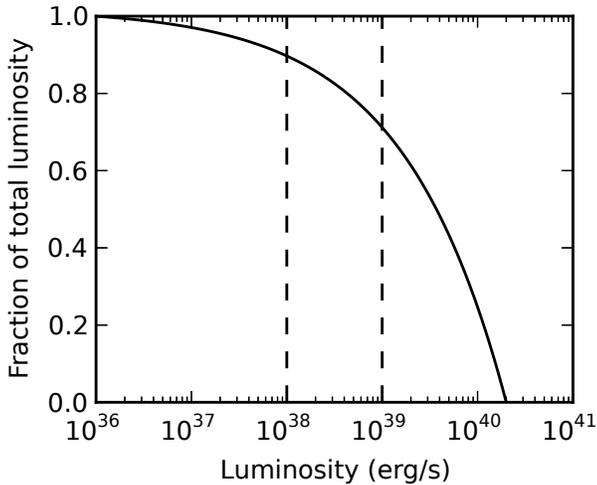}}
\caption{Fraction of total X-ray luminosity above a threshold luminosity for an X-ray luminosity function with a slope of $-1.6$.}
\label{xlf} \end{figure}

\subsection{X-ray spectra}
\label{sec:spectra}

It is clear that the number of HMXBs and their total X-ray luminosity is enhanced at low metallicity.  A key question for understanding the contribution of X-ray binaries to heating and reionization is the shape of their emission spectrum.  The HMXB X-ray luminosity function (XLF) is well described by a power-law distribution with an exponent near -1.6 and a cut-off near $10^{41} \rm \, erg \, s^{-1}$ \citep{Grimm03}.  The XLF in low metallicity galaxies has the same form and exponent \citep{Brorby14}.  Figure~\ref{xlf} shows the integrated luminosity from sources with luminosities above a given threshold.  The shallow slope of the HMXB XLF implies that the total luminosity is dominated by bright sources.  Half of the flux from active sources (with $L_X > 10^{36} \rm \, erg \, s^{-1}$) is produced by sources with $L_X > 4 \times 10^{39} \rm \, erg \, s^{-1}$, 71 percent by sources with $L_X > 10^{39} \rm \, erg \, s^{-1}$, and 90 per cent by sources with $L_X > 10^{38} \rm \, erg \, s^{-1}$.

The total luminosity and, hence, the spectral shape of the emission from a population of HMXBs is dominated by ULXs and HMXBs in high luminosity states.  Black hole X-ray binaries in the Milky Way exhibit hard spectra, following a power-law form with photon index near 1.7, at low luminosities, and then transition to states with softer spectra at higher luminosities.  These softer spectra are a combination of a thermal component with temperatures up to a few keV and steep power-laws with photon indices near 2.5.  These sources reach luminosities up to $\sim 10^{39} \rm \, erg \, s^{-1}$\citep{Remillard06}.

ULXs produce a large fraction of the total X-ray luminosity and exhibit spectra that show distinct curvature in the 2--8~keV band.  This was first demonstrated with high statistical quality XMM-Newton spectra \citep{Stobbart06,Gladstone09}.  The curvature in ULX spectra has recently been confirmed with Suzaku and joint NuSTAR/XMM-Newton observations.  Suzaku observations of the starburst galaxy M82 show that its X-ray spectrum, seen in Chandra imaging to be dominated by a ULX \citep{Kaaret01}, has a cut-off energy of $5.7 \pm 0.7$~keV when fitted with an exponentially cut-off model \citep{Miyawaki09}.  XMM-Newton/NuSTAR spectra of the ULX NGC 1313 X-2 show cut-off energies near 2.5~keV while those of NGC 1313 X-1 show cut-off energies in the range 5--9~keV \citep{Bachetti13}.  The HMXB in the very low metallicity galaxy I Zw 18 ($Z/Z_{\odot} = 0.019$), shows a transition from a hard power-law spectum at low luminosity to a curved spectrum at high luminosity \citep{Kaaret13}.  When fitted with an exponentially cut-off model, the high luminosity spectrum has a cut-off energy of $2.1 \pm 0.6$~keV and a photon index of $0.80 \pm 0.26$.


Some recent papers, e.g.\ \citet{McQuinn12}, have assumed a hard power-law with photon index of 1.7 for the HMXB spectrum based on measurements of HMXBs in nearby galaxies made by \citet{Swartz04} using Chandra.  These results are influenced by the low numbers of counts in the Chandra spectra.  \citet{Swartz04} find that a power law is unacceptable for 39 per cent of the spectra with more than 1000 counts but only 8 per cent of the spectra with fewer than 200 counts.  The average photon index was derived using only those spectra for which a power law provided a statistically acceptable fit, systematically excluding those spectra exhibiting evidence for curvature and indirectly excluding a higher fraction of spectra with larger numbers of counts.  Other work has considered the transition to softer spectra at the luminosities seen from Milky Way X-ray binaries, but not the curved spectra seen from ULXs \citep{Fragos13,Power13}.  The higher quality XMM-Newton, NuSTAR, and Suzaku spectra provide a better measurement of the true spectral shape of the very bright HMXBs that dominate the total HMXB luminosity and consistently show curvature at high energies.


\begin{figure}
\centerline{\includegraphics[width=3.2in,angle=0]{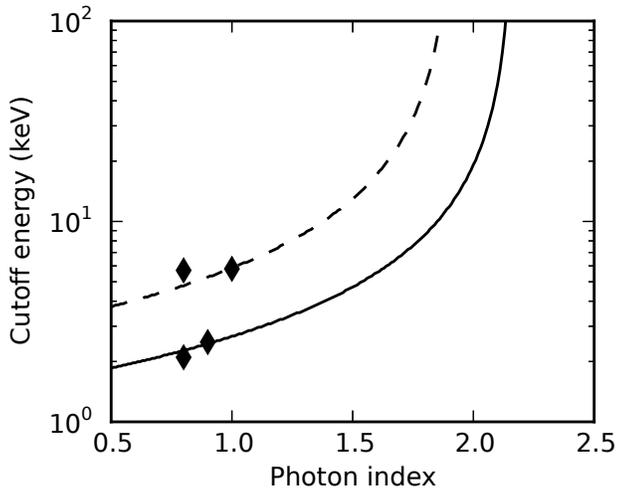}}
\caption{Photon index versus cut-off energy for an exponentially cut-off power-law model consistent with the ratio of Chandra counts in the 2--8/0.5--2~keV bands.  The solid line corresponds to the ratio measured by \citet{Cowie12} for $z=0-1$ and the dashed line for $z=1-2$.  The diamonds show the parameters measured for various ULXs: I Zw 18 X-1, M82 X-1, NGC 1313 X-1, and NGC 1313 X-2.}
\label{cowie_cutpl} \end{figure}

It is possible to get some constraints on the spectral shape of the X-ray emission from HMXBs at moderate redshifts from deep X-ray surveys.  \citet{Cowie12} performed a stacking analysis of star-forming galaxies in the Chandra Deep Field South and calculated the counts in two energy bands (0.5--2~keV and 2--8~keV) for sets of galaxies in different redshift intervals.  The galaxies are significantly detected in the harder band only for the redshift bands $z=0-1$ and $z = 1-2$.  The hardness ratios are $0.28 \pm 0.08$ and $0.36 \pm 0.11$, respectively.  Using the known response of Chandra it is possible to calculate the expected counts ratio for an exponentially cut-off power-law model with a given cut-off energy and photon index -- note that it is necessary to take into account the redshift in order to interpret the cut-off energy in the galaxy rest frame.  Figure~\ref{cowie_cutpl} shows the exponentially cut-off power law model parameters consistent with the measured counts ratios for the $z=0-1$ and $z = 1-2$ redshift bands.  The allowed parameter range is consistent with the parameters measured for the ULXs mentioned above.

\begin{figure}
\centerline{\includegraphics[width=3.2in,angle=0]{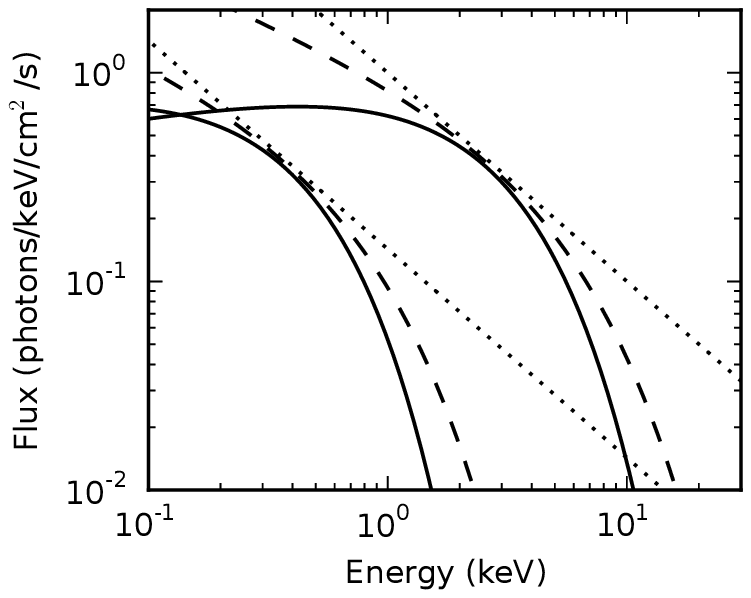}}
\centerline{\includegraphics[width=3.2in,angle=0]{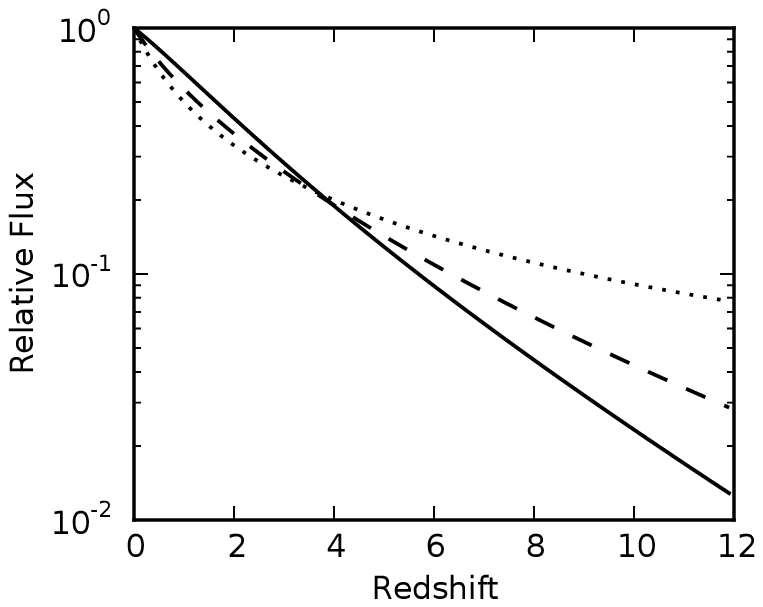}}
\caption{Top: various spectral models at $z=0$ (upper curves) and $z=6$ (lower curves).  The relative flux normalizations account for the K-correction, but not distance.  Bottom: flux in the 0.5--2~keV band for various spectral models as a function of redshift normalized to the flux at $z=0$.  In both panels, the solid line shows an exponentially cut-off power law with a cut-off energy of 2.1~keV and a photon index of 0.8, the dashed line shows the same model with a cut-off energy of 6.0~keV and and photon index of 1.5, and the dotted line shows a power law with a photon index of 1.7.}
\label{zspec} \end{figure}

\begin{figure*}
\centerline{\includegraphics[width=6.25in,angle=0]{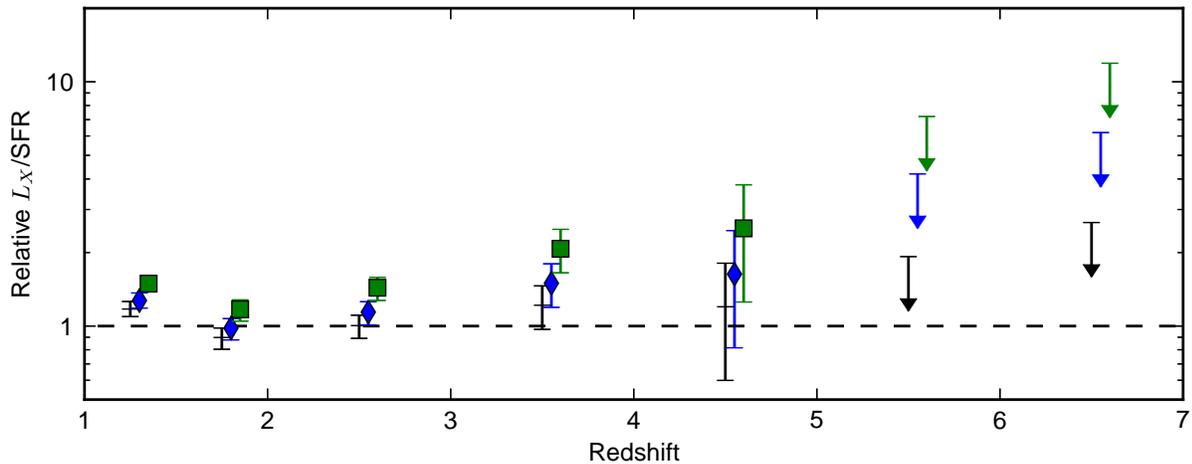}}
\caption{Ratio of X-ray luminosity to star formation rate, $L_X$/SFR, normalized to the value of $2.6 \times 10^{39} \rm \, erg \, s^{-1} \, M_{\odot}^{-1} yr$ measured in nearby, near solar metallicity galaxies) for various spectral models.  The crosses (black) are for a power law with photon index of 2.  The diamonds (blue) are for a cut-off power law with a cut-off energy of 6.0~keV and a photon index of 1.5.  The squares (green) are for an exponentially cut-off power law with a cut-off energy of 2.1~keV and a photon index of 0.8.  The latter two models are shifted along the x-axis by 0.05 and 0.1, respectively, for clarity.  Error bars are 68 per cent confidence intervals while upper limits are 90 per cent confidence.}
\label{lxsfrz} \end{figure*}

\section{Implications}
\label{sec:implications}

Studies of galaxies with low metallicity suggest that the total luminosity of early X-ray binaries (normalised to SFR) is likely enhanced relative to near-solar metallicity systems and that the total spectrum is likely dominated by systems in high luminosity states with curved spectra.  This spectral curvature may potentially affect: constraints on early X-rays from measurements of the soft X-ray background, constraints on the ratio of X-ray luminosity to SFR at high redshift obtained from deep X-ray surveys, the detectability of the first HMXBs, the morphology of X-ray heating, and the 21~cm radio signal from the epoch of reionization.

Figure~\ref{zspec} shows power law and exponentially cut-off power law spectra at redshifts of $z=0$ and $z=6$ in the top panel and the observed flux in the 0.5--2~keV band as a function of redshift for these models in the bottom panel -- essentially the `K-correction' for each spectral model.  This band is often used in deep X-ray surveys.  For a fixed observational energy band, an exponential cut-off in the spectrum has a strong effect on the observed flux as the red-shifted cut-off energy moves into the band.  Because of the rather hard photon index measured for I Zw 18, this effect is somewhat negated at low redshifts because that spectrum rises with energy at energies below the cut-off.

Spectral curvature affects observational constraints on X-rays from early binaries.  With the assumption of a hard, power law spectrum for the HMXB emission, a significant HMXB contribution to reionization is excluded \citep{McQuinn12}.  However, even with a break energy as high as 5~keV, the cut-off energy of emission from $z = 10$ would be redshifted to 0.5~keV, below the 0.5--2 keV band typically used in deep X-ray surveys.  If the total HMXB emission spectrum has significant curvature as seen in the ULX spectra, then constraints from the soft X-ray background on the emission from early X-ray binaries become very weak.  X-ray emission sufficient to fully power reionization is allowed \citep{Power13}.

Deep X-ray surveys can also be used to constrain the emission from HMXBs and the ratio of X-ray luminosity to SFR ($L_X$/SFR) at high redshift.  As noted above, \citet{Cowie12} performed a stacking analysis of star-forming galaxies in the Chandra Deep Field South.  In addition to calculating the total X-ray emission in various redshift bands, they also estimated the total SFR of the target galaxies in each band.  Correcting for the effect of redshift on the observed flux in the 0.5--2~keV band using a power-law model with a photon index of 2, they concluded there is no evidence for redshift evolution of $L_X$/SFR.  In contrast, Basu-Zych et al. (2013a) found an increase in $L_X$/SFR at high redshifts.

Figure~\ref{lxsfrz} shows the effect of spectral curvature on the redshift dependent estimates of $L_X$/SFR of Cowie et al. (2012.  The increase in the K-correction caused by the spectral curvature progressively increases the estimated values of $L_X$/SFR at increasing redshifts.  The results then suggest a moderate increase in $L_X$/SFR at intermediate redshifts, $z=3-5$, and are consistent with relatively large enhancements at high redshifts.  For the exponentially cut-off power law spectrum with a cut-off energy of 5~keV, the upper limits (90 per cent confidence) on the enhancement in $L_X$/SFR are 4.2 and 6.2 in the two highest redshift bins, $z=$5--6 and 6--7.  With a cut-off energy of 2.1~keV, the corresponding upper limits are 7.2 and 11.9.  The latter constraints are consistent with the increase in $L_X$/SFR expected due to the decrease in metallicity at high redshifts while the former would suggest that the results from low-metallicity nearby galaxies, mainly blue compact dwarfs in the samples of \citet{Prestwich13} and \citet{Brorby14}, may overpredict the X-ray emission from the larger galaxies detected at high redshifts.

\begin{figure}
\centerline{\includegraphics[width=3.2in,angle=0]{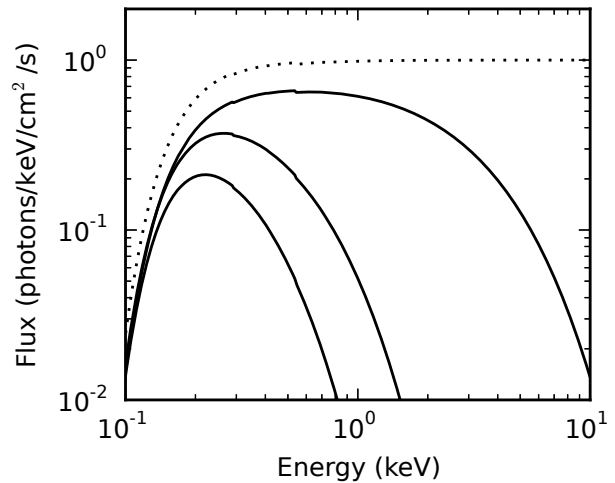}}
\caption{Red-shifted spectra with the effect of interstellar absorption.  The solid curves, from top to bottom, show the spectra at $z=0$, 6, and 12.  The relative flux normalizations account for the K-correction, but not distance. The dotted curve is the fractional absorption.}
\label{lockman} \end{figure}

Spectral curvature combined with interstellar absorption within the Milky Way may limit the detectability of HMXBs at high redshifts.  The lines of sight with the lowest column density within the Milky Way are within the Lockman hole \citep{lockman_hole} and have absorption column densities near $N_H = 6 \times 10^{19} \rm \, cm^{-2}$.  Figure~\ref{lockman} shows the X-ray absorption for this column density along with an exponentially cut-off power law spectrum with cut-off energy of 2.1~keV at redshifts of $z=0$, 6, and 12.  It is apparent that interstellar absorption has a severe effect, strongly reducing the X-ray flux and the band over which X-rays are detected.  We note that current CCD detectors operate efficiently only above energies of about 0.2~keV.  This may place an even more severe constraint on the detectability of high redshift HMXBs.

Spectral curvature will also affect how far X-rays from early HMXBs penetrate into the IGM and therefore the morphology of X-ray heating.  This affects the 21-cm radio emission from the epoch of reionization and it may be possible to place constraints on the total spectrum of early X-ray emission when adequate 21-cm maps become available \citep{Pritchard07}.  For truly pristine gas with zero metallicity, the highest ionisation edge is at 54.4~eV so the effect of spectral curvature above 2~keV is not so pronounced.  However, metal enrichment local to the site of HXMB formation could have an effect.  Spectral curvature will also have an effect on the temporal extent of X-ray heating since it affects how the observed/absorbed spectrum evolved with redshift.

\vspace{-12pt}
\section{Summary}
\label{sec:summary}

Study of nearby high mass X-ray binaries can provide insight into HMXBs formed early in the history of the universe.  The dominant factor leading to change in the properties of early HMXBs compared with current HMXBs is the reduced metallicity in the early universe.  Multiple observational and theoretical studies have shown that X-ray binary production is enhanced at low metallicity in terms of the number of HMXBs produced for a given amount of star formation, particularly for HMXBs with high luminosities, and the total X-ray luminosity produced.

The total luminosity of current populations of HMXBs is dominated by sources at high luminosities.  These sources exhibit X-ray spectra that show curvature above 2~keV.  The same is likely true of HMXB populations at high redshifts.  The spectral curvature changes the K-correction for X-rays from HMXBs in a manner that greatly weakens the constraints on X-ray emission of early HMXBs obtained from the soft X-ray background.  Applied to deep X-ray surveys of star forming galaxies, the modified K-correction suggests a moderate increase in the ratio of X-ray luminosity to SFR ($L_X$/SFR) at intermediate redshifts, $z=3-5$, and is not inconsistent with a large enhancement at high redshifts, $z=6-7$.



\label{lastpage}


\begin{thebibliography}{99}

\bibitem[Bachetti et al.(2013)]{Bachetti13} Bachetti, M., Rana, V., Walton, D.J., et al.\ 2013, ApJ, 778, 163

\bibitem[Basu-Zych et al.(2013a)]{Basu13a} Basu-Zych, A.R., Lehmer, B.D., Hornschemeier, A.E., et al.\ 2013a, ApJ, 762, 45

\bibitem[Basu-Zych et al.(2013b)]{Basu13b} Basu-Zych, A.R., Lehmer, B.D., Hornschemeier, A.E., et al.\ 2013b, ApJ, 774, 152

\bibitem[Belczynski et al.(2010)]{Belczynski10} Belczynski, K.\ et al.\ 2010, ApJ, 714, 1217

\bibitem[Bromm(2013)]{Bromm13} Bromm, V.\ 2013, Rep.\ Prog. Phys., 76, 112901


\bibitem[Brorby et al.(2014)]{Brorby14} Brorby, M., Kaaret, P., Prestwich, A.\ 2013, MNRAS, submitted

\bibitem[Cowie et al.(2012)]{Cowie12} Cowie, L.L., Barger, A.J., Hasinger, G.\ 2012, ApJ, 748, 50

\bibitem[David et al.(1992)]{David92} David, L.P., Jones, C., \& Forman, W. 1992, ApJ, 388, 82

\bibitem[Dray(2006)]{Dray06} Dray, L.M.\ 2006, MNRAS, 370, 2079

\bibitem[Fragos et al.(2013)]{Fragos13} Fragos, T., Lehmer, B., Tremmel, M., et al.\ 2013, ApJ, 764, 41

\bibitem[Fryer(1999)]{Fryer99} Fryer, C.L.\ 1999, ApJ, 522, 413

\bibitem[Fryer \& Kalogera(2001)]{Fryer01} Fryer, C.L., Kalogera, V.\ 2001, ApJ, 554, 548

\bibitem[Gladstone et al.(2009)]{Gladstone09} Gladstone, J.C.,  Roberts, T.P., Done, C.\ 2009, MNRAS, 397, 1836

\bibitem[Griffiths et al.(1990)]{Griffiths90} Griffiths, R.E.,\& Padovani, P. 1990, ApJ, 360, 483

\bibitem[Grimm et al.(2003)]{Grimm03} Grimm, H.-J., Gilfanov, M., \& Sunyaev, R. 2003, MNRAS, 339, 793

\bibitem[Heger et al.(2003)]{Heger03} Heger, A., Fryer, C.L., Woosley, S.E., Langer, N., Hartmann, D.H.\ 2003, ApJ, 591, 288

\bibitem[Irwin et al.(2004)]{Irwin04} Irwin, J.A., Bregman, J.N., Athey, A.E.\ 2004, ApJ, 601, L143

\bibitem[Jeon et al.(2013)]{Jeon13} Jeon, M., Pawlik, A.H., Bromm, V., Milosavljevi\'c, M.\ 2013, MNRAS submitted, arXiv:1310.7944

\bibitem[Kaaret et al.(2001)]{Kaaret01} Kaaret, P.\ et al.\ 2001, MNRAS, 321, L29

\bibitem[Kaaret \& Alonso-Herrero(2008)]{Kaaret08} Kaaret, P.\ \& Alonso-Herrero, A.\ 2008, ApJ, 682, 1020

\bibitem[Kaaret et al.(2011)]{Kaaret11} Kaaret, P.\, Schmitt, J., \& Gorski, M.\ 2011, ApJ, 741, 10

\bibitem[Kaaret \& Feng(2013)]{Kaaret13} Kaaret, P.\ \& Feng, H.\ 2013, ApJ, 770, 20

\bibitem[Linden et al.(2010)]{Linden10} Linden, T.\ et al.\ 2010, ApJ, 725, 1984

\bibitem[Lockman et al.(1986)]{lockman_hole} Lockman, F.J., Jahoda, K., McCammon, D.\ 1986, ApJ, 302, L432

\bibitem[Loeb(2010)]{Loeb10} Loeb, A.\ 2010, How did the first stars and galaxies form?, (Princeton University Press)

\bibitem[Madau \& Efstathiou(1999)]{Madau99} Madau, P.\ \& Efstathiou, G.\ 1999, ApJ, 517, L9

\bibitem[Mapelli et al.(2009)]{Mapelli09} Mapelli, M.\ et al.\ 2009, MNRAS, 395, L71

\bibitem[Mapelli et al.(2010)]{Mapelli10} Mapelli, M.\ et al.\ 2010, MNRAS, 408, 234

\bibitem[McQuinn et al.(2012)]{McQuinn12} McQuinn, M.\ 2012, MNRAS, 426, 1349

\bibitem[Miyawaki et al.(2009)]{Miyawaki09} Miyawaki, R., Makishima, K., Yamada, S., Gandhi, P., Mizuno, T., Kubota, A., Tsuru, T.G., Matsumoto, H.\ 2009, PASJ, 61, 263

\bibitem[Mirabel et al.(2011)]{Mirabel11} Mirabel, I.F., Dijkstra, M., Laurent, P., Loeb, A., Pritchard, J.R.\ 2011, A\&A, 528, A149


\bibitem[Power et al.(2013)]{Power13} Power, C., James, G., Combet, C., Wynn, G.\ 2013, ApJ, 764, 76

\bibitem[Prestwich et al.(2013)]{Prestwich13} Prestwich, A.H., Tsantaki, M., Zezas, A., et al.\ 2013, ApJ, 769, 92

\bibitem[Pritchard \& Furlanetto(2007)]{Pritchard07} Pritchard, J.R.\ \& Furlanetto, S.R.\ 2007, MNRAS, 376, 1680

\bibitem[Ranalli et al.(2003)]{Ranalli03} Ranalli, P., Comastri, A., \& Setti, G.\ 2003, A\&A, 399, 39

\bibitem[Remillard \& McClintock (2006)]{Remillard06} Remillard, R.E.\ \& McClintock, J.E.\ 2006, ARA\&A, 44, 49

\bibitem[Ricotti \& Ostriker(2004)]{Ricotti04} Ricotti M., Ostriker J. P., 2004, MNRAS, 352, 547

\bibitem[Salvaterra et al.(2012)]{Salvaterra12} Salvaterra, R., Haardt, F., Volonteri, M., Moretti, A.\ 2012, A\&A, 545, L6

\bibitem[Swartz et al.(2004)]{Swartz04} Swartz, D.A., Ghosh, K.K., Tennant, A.F., Wu, K.\ 2004, ApJS, 154, 519

\bibitem[Stobbart et al.(2006)]{Stobbart06} Stobbart, A.-M., Roberts, T.P., Wilms, J.\ 2006, MNRAS, 368, 397

\bibitem[Treister et al.(2013)]{Treister13} Treister, E., Schawinski, K., Volonteri, M., Natarajan, P.\ 2013, ApJ, 778, 130

\bibitem[Venkatesan et al.(2001)]{Venkatesan01} Venkatesan, A., Giroux M.L., Shull J.M., 2001, ApJ, 563, 1

\bibitem[Wheeler \& Johnson(2011)]{Wheeler11} Wheeler, J.C., Johnson, V.\ 2011, ApJ, 738, 163


\bibitem[Zampieri \& Roberts(2009)]{Zampieri09} Zampieri, L., Roberts, T.P.\ 2009, MNRAS, 400, 677

\end{thebibliography}
\end{document}